\documentstyle[12pt]{article}

\def\noin{\noindent}
\def\non{\nonumber}
\def\beq{\begin{equation}}
\def\eeq{\end{equation}}
\def\ben{\begin{eqnarray}}
\def\een{\end{eqnarray}}
\def\lab{\label}

\def\ra{\rightarrow}
\def\l{\left}
\def\r{\right}
\def\bm{\boldmath}
\def\del{\partial}

\def\cH{{\cal H}}
\def\cL{{\cal L}}
\def\cO{{\cal O}}

\def\gSp{\sigma_{\pi N}}
\def\mpi{m_\pi}

\def\Nb{\bar{N}}
\def\uB{\bar{u}}
\def\dB{\bar{d}}

\def\TpN{T_{\pi N}}

\def\half{\frac{1}{2}}

\def\th{\theta}

\def\mN{m_{\mbox{\tiny{$N$}}}}
\def\psiN{\psi_{\mbox{\tiny{$N$}}}}
\def\psibarN{\bar{\psi}_{\mbox{\tiny{$N$}}}}

\def\bfk{{\bf k}}
\def\bfp{{\bf p}}
\def\bfpi{\mbox{\boldmath$\pi$}}
\def\bftau{\mbox{\boldmath$\tau$}}
\def\bfgam{\mbox{\boldmath$\gamma$}}

\def\PL{{\it Phys. Lett.\,\,}}
\def\NP{{\it Nucl. Phys.\,\,}}
\def\PR{{\it Phys. Rev.\,\,}}

\def\thefootnote{\fnsymbol{footnote}}

\begin{document}

\noindent{USC(NT)--94--5}
\vskip 1cm
\begin{center}
{\bf CHIRAL PERTURBATION
IN DENSE MATTER AND}\\
{\bf MESON CONDENSATION CONTROVERSY\footnote{
Invited talk at the International Symposium on
Frontiers in Nuclear Structure Physics,
RIKEN, Saitama, Japan; March 1994}
}\\

\vspace{0.8cm}
Kuniharu Kubodera\footnote{
Supported in part
by the NSF under Grant No. PHYS-9310124.}\\
{\it Department of Physics and Astronomy,
University of South Carolina}\\
{\it Columbia, South Carolina 29208, USA}
\end{center}

\vspace{0.8cm}
\centerline{ABSTRACT}
\begin{quote}
\noin
\footnotesize{An outstanding problem
in the study of possible kaon condensation is the
glaring discrepancy between the results of
chiral perturbation theory and
those of the PCAC-plus-current-algebra approach.
I discuss here what causes this discrepancy
and what needs to be done to solve the problem.
In addition, I point out the importance of
examining the validity of the non-relativistic approximation
$<\!\!\bar{\psi}\psi\!\!>\;\sim\;
<\!\!\bar{\psi}\gamma_0\psi\!\!>$,
an approximation that is universally employed in the
existing treatments of kaon condensation.}
\end{quote}

\vspace{0.4cm}
\renewcommand\thefootnote{\arabic{footnote}}
\setcounter{footnote}{0}
Since the seminal work
by Kaplan and Nelson$^{1,2}$,
kaon condensation
and its possible astrophysical consequences
have been a subject of intensive studies$^{3-12}$.
Our interest in this phenomenon
has received another strong boost
from Brown and Bethe's recent suggestion$^{13}$
that the significant softening of the equation of state
due to kaon condensation can drastically change
the stellar collapse scenario, providing a possible solution
to the long-standing ``association problem"
-- the puzzling paucity of neutron stars
as compared with supernova events.
In my view, there are at least two major issues
that need to be investigated before a solid conclusion
on kaon condensation can be drawn.
The first is that theoretical predictions
vary drastically according to whether
one uses chiral perturbation theory (ChPT)
or the current-algebra-plus-PCAC approach.
The second point is that all the existing treatments rely
on the non-relativistic approximation
$\rho_s\sim\rho$,
where $\rho_s=\,<\!\!\bar{\psi}\psi\!\!>$ is
the Lorentz-scalar density of baryons and
$\rho=\,<\!\!\bar{\psi}\gamma_0\psi\!\!>$ is the
baryonic density.
In this talk I shall address these two issues.

In the ``standard" approach based on chiral perturbation$^{1-12}$,
the attractive force that drives condensation
is provided primarily by the $K$-$N$ sigma term, and,
for typical values of $\Sigma_{KN}$,
the critical density for kaon condensation
is predicted to be $\rho_c = \,2 \sim 3 \, \rho_0$
($\rho_0$ = normal nuclear density).
However, the validity of this approach
has been questioned by several authors$^{14-18}$.
According to Yabu et al.\ $\!\!^{13,14}$,
the use of $K$-$N$ scattering amplitudes
that respect the current algebra and PCAC
does not lead to kaon condensation.
To clarify the origin of this discrepancy
is a crucial step in pinning down
the elusive issue of kaon condensation.
Very recently, Yabu, Myhrer and I myself
have shown$^{15}$ that, in order to resolve
this serious disagreement,
one needs to include in the starting Lagrangian itself
terms that are of higher order in density
than previously considered.
The following is a brief account
of this latest work$^{15}$.

Since the sigma term is the central issue,
let us focus our attention on
the role of the sigma term.
Furthermore, for the illustrative purpose,
let us consider s-wave pion condensation
rather than kaon condensation itself.
To set the stage, we first summarize the basic feature of
the original treatment of meson condensation
based on a chiral effective Lagrangian.
As a toy model
we work with the lowest-order ChPT expansion
and discard all terms that are not of direct relevance
to our argument.
Thus, we use
\beq
\cL_{1} = \frac{1}{2} \l[ -\phi (\Box+\mpi^2) \phi
+ \frac{\gSp}{f^2} \phi^2 {\Nb N} \r], \lab{eQa}
\eeq
where $\phi(x)$ and $N(x)$
are the pion and the nucleon field, respectively,
and $f$ is the pion decay constant;
$\gSp \equiv
\frac{1}{2}(m_u+m_d) \mbox{$<N|{\uB u}+{\dB d}|N>$}$
is the $\pi N$ sigma term.
For $\cL_{1}$,
the $\pi$-$N$ scattering amplitude in tree approximation
is given simply by
\beq
\TpN^{(1)} = \frac{\gSp}{f^2}. \lab{eQb}
\eeq
To estimate the effective pion mass $\mpi^*$ in nuclear matter,
we follow a common practice in the literature
and use the mean-field approximation, $\Nb N \sim \rho$.
Then the pion dispersion relation reads
$\omega^2-\bfk^2-\mpi^2+{\rho\gSp / f^2} =0$.
The effective pion mass $\mpi^*$ is defined
by $\mpi^* \equiv \omega$($\bfk=0$),
and the critical density $\rho_c$
for pion condensation is determined from the condition
$m_\pi^*=0$.
For the case at hand we obtain
\beq
[\mpi^*(1)]^2 =\mpi^2 -\rho \frac{\gSp}{f^2}. \lab{eQc}
\eeq
and so
\beq
\rho_c =\frac{\mpi^2 f^2}{\gSp}. \lab{eQd}
\eeq

Another approach employed by Yabu et al.\ $\!\!^{13,14}$ is based on the PCAC
an
The pion extrapolating field $\pi(x)$ in this approach is
taken to be
\beq
\pi(x) \equiv \frac{1}{\mpi^2 f} \del_\mu A^\mu(x),
\lab{eQe}
\eeq
where $A^\mu(x)$ is the axial current.
With this choice of $\pi(x)$, the $\pi$-$N$ scattering amplitudes
for on- and off-shell momenta of the pions are ``defined'' by
\beq
     \TpN^{(2)} =i^2 (\mpi^2-(k')^2) (\mpi^2-k^2)
  {\int d^4x d^4y\,} e^{ik'x}e^{-iky}<\!\!N'| T \pi(x) \pi(y) |N\!\!>,
\lab{eQf}
\eeq
where $k$ ($k'$) is the incoming (outgoing) pion momentum.
The amplitude $\TpN^{(2)}$ fulfils
the Adler condition and, at the Weinberg point,
it also satisfies the well-known relation
with the sigma term.
For forward scattering,
the general form of $\TpN$
that is consistent with the low-energy theorems
can be written as
\beq
\TpN^{(2)} =
{k^2 +(k')^2-\mpi^2 \over f^2 \mpi^2} \gSp +\TpN',
\lab{eQg}
\eeq
where only the $\gSp$-dependent terms are explicitly shown;
these terms become identical to the amplitude
in eq.(\ref{eQb}) for on-mass-shell mesons.
The remaining term, $\TpN'$, contains the Born terms,
the Weinberg-Tomozawa term, etc.,
and plays an important role in a realistic
calculation of the on-shell $\pi$-$N$ scattering
amplitude$^{14,16-18}$.
However, we neglect these terms here
to delineate the role of the sigma term.
In the mean field approximation,
$\mpi^*$ that corresponds
to the $\pi$-$N$ amplitude in eq.(\ref{eQg}) is given by$^{16}$
\beq
[\mpi^*(2)]^2 =\mpi^2 \,\l(1 +\rho\,\frac{\gSp}{\mpi^2f^2}\r) \;
\l(1+2 \rho\,\frac{\gSp}{\mpi^2f^2}\,\r)^{-1}\,\,.
\lab{eQi}
\eeq

Although $\mpi^*(1)$ and $\mpi^*(2)$
coincide with each other for sufficiently low densities,
they behave very differently for large values of $\rho$.
In particular, $\mpi^*(2) \ra \mpi/\sqrt{2}$ as
$\rho \ra \infty$,
a feature that makes meson condensation highly unlikely.
The difference between $\mpi^*$ of eq.(\ref{eQi})
and $\mpi^*$ of eq.(\ref{eQc}) summarizes
the basic problem involved
in the existing treatments of meson condensation.
In view of the well-established phenomenological success
of ChPT and the PCAC approach,
it is very surprising that their predictions on $\mpi^*$\,,
as they stand, differ so drastically.
We of course know that,
by working with a chiral effective Lagrangian
with gauged external source terms,
it is possible to recover the Adler condition in ChPT$^{19}$.
This procedure, however, amounts to using
$\pi$ of eq.(6) as the pion field operator
instead of the original field
that appears in the effective Lagrangian.
Therefore, this ``reconciliation" between ChPT and PCAC
does not solve the difficulty we are encountering here.

Now, the two amplitudes,
$\TpN^{(1)}$ and $\TpN^{(2)}$ ,
although identical on the mass shell,
exhibit totally different off-mass-shell behaviors.
As is well known,
the off-mass-shell values of
the $\pi$-$N$ amplitudes depend on
the choice of an extrapolating field.
In the present case the difference
between $\TpN^{(1)}$ and $\TpN^{(2)}$
reflects the two non-equivalent extrapolating fields,
$\phi(x)$ [eq.(\ref{eQa})] and
$\pi(x)$ [eq.(\ref{eQe})].
One might be tempted to ascribe the
variance between $\mpi^*(1)$ and $\mpi^*(2)$
to the different off-mass-shell behaviors of the $\pi$-$N$
scattering amplitudes.
This interpretation, however, is not tenable
for the reason given immediately below.
For a given Lagrangian $\cL$, the finite-density
pion Green function is defined by
\beq
G_{\rho}(x;\varphi) \,=\,
<\!\rho| \,T \varphi(x) \varphi(0) \,|\rho\!>, \lab{eQla}
\eeq
where $|\rho\!\!>$ is the ground state
(with baryon density $\rho$)
of the system governed by $\cL$,
and $\varphi(x)$ is an arbitrary operator
for the pion field.
One can use any field $\varphi$ so long as
it connects one-pion state to vacuum, i.e.,
$<\!\!\pi|\varphi(x)|0\!\!> \;\neq 0$.
The energy $E_n$
of a pionic-mode intermediate state $|n\!\!>$
that can be connected to $|\rho\!\!>$ via $\varphi$
gives the pole position of $G_{\rho}(x;\varphi)$.
Since $E_n$ is determined by $\cL$ alone,
it is {\it independent}
of the choice of $\varphi$.
This means that $\mpi^*$,
which is uniquely given
by the pole position of $G_{\rho}(x;\varphi)$,
must also be independent of $\varphi$.
One therefore cannot attribute
the discrepancy between $\mpi^*(1)$ and $\mpi^*(2)$
to the off-mass-shell problem.

To obtain a better understanding of the true nature
of the problem,
let us consider an effective Lagrangian $\cL_2$
which, at the tree level, reproduces
the first term of the $\pi$-$N$ scattering amplitude eq.(\ref{eQg})
and hence leads to the effective mass eq.(\ref{eQi}):
\beq
\cL_{2} =\frac{1}{2} \l[ -\pi (\Box+\mpi^2) \pi
 -\frac{\gSp}{f^2}(\pi^2 +\frac{2}{\mpi^2}\pi \Box \pi)
{\Nb N} \r]. \lab{eQh}
\eeq
$\cL_2$ differs from $\cL_1$
by the existence of the interaction term that involves
$\Box\pi$ (``box term'').
In ChPT, the box terms in the {\it pure mesonic sector}
can be eliminated by redefining the meson field.
Since the meson field in ChPT
is nothing more than an integration variable and
has no physical meaning by itself,
we need to examine
to what extent the difference
between $\cL_1$ and $\cL_2$ can be transformed away
via a meson-field redefinition.
To this end, we apply the mean field approximation,
${\Nb N} \rightarrow \rho$,
to eq.(\ref{eQh})\footnote{This also allows us to circumvent
mathematical subtleties
associated with the treatment of operator products.}
and introduce a new meson field $\tilde{\phi}(x)$
defined by
\beq
\pi(x)=
\l( 1 -\rho \frac{\gSp}{f^2\mpi^2} \r) \tilde{\phi}(x).
\lab{eQj}
\eeq
In terms of $\tilde{\phi}(x)$, the Lagrangian (\ref{eQh})
can be recast into
\beq
    \cL_2 =\l( 1 -\rho \frac{\gSp}{f^2\mpi^2} \r)^2
\frac{1}{2} \l[ -\tilde{\phi}\,(\Box+\mpi^2) \tilde{\phi}
-\frac{\gSp}{f^2}\,(\tilde{\phi}^2 + \frac{2}{\mpi^2}\,
\tilde{\phi}\,\Box \tilde{\phi}\,)\,\rho \r]. \lab{eQk}
\eeq
Expansion in $\rho$ gives
\beq
    \cL_2 =\frac{1}{2}
    \l[ -\tilde{\phi}\,(\Box+\mpi^2) \tilde{\phi}
+ \rho\,\frac{\gSp}{f^2}\,\tilde{\phi}^2 \r] +\cO(\rho^2).
\lab{eQl}
\eeq
The transformed  Lagrangian (\ref{eQl}) is
identical to eq.(\ref{eQa}),
if the terms of $\cO(\rho^2)$ are neglected,
and this is consistent with the fact
that $\mpi^*(1) =\mpi^*(2)$ when we ignore
terms of $\cO(\rho^2)$

At the $\cO(\rho^2)$ level,
$\cL_1$ is no longer identical to $\cL_2$,
and this feature is responsible for the difference between
$\mpi^*(1)$ and $\mpi^*(2)$.
Although this statement itself is correct,
the real significance of this statement
hinges upon the following crucial question:
Do the existing formalisms allow us to
make a meaningful distinction
between $\mpi^*(1)$ and $\mpi^*(2)$ ?
For the sake of clarity, we rephrase this question
in terms of $G_{\rho}(x;\varphi)$, eq.(\ref{eQla}).
For the Lagrangian $\cL_2$,
one may consider two Green functions,
$G^{(2)}(x;\pi)  \equiv  G_{\rho}(x;\varphi=\pi)$
and $ G^{(2)}(x;\tilde{\phi}) \equiv
G_{\rho}(x;\varphi=
[1 -(\rho \gSp /f^2\mpi^2)] \tilde{\phi})$.
Although these Green functions are {\it not\ } identical,
their pole positions give the same effective mass
$\mpi^*(2)$, eq.(\ref{eQi}).
On the other hand, if we consider the Green function
$G^{(1)}(x;\phi) \equiv G_{\rho}(x;\varphi=\phi)$
governed by $\cL^{(1)}$,
with $\phi$ being the field appearing in eq.(\ref{eQa}),
the pole position will move to
$\mpi^*(1)$, eq.(\ref{eQc}), reflecting a change
in the basic Lagrangian\footnote{
One could also work with $G^{(1)}(x;\pi)$ using eq.(\ref{eQj}),
which would give the same effective mass as $G^{(1)}(x;\phi)$.}.
Now, if there is a unique criterion to decide which
effective Lagrangian, $\cL^{(1)}$ or $\cL^{(2)}$,
describes reality better,
then one would know which effective mass to use,
$\mpi^*(1)$ or $\mpi^*(2)$.
So, the crucial question is whether
the formalisms so far developed allow us to
decide which of $\cL^{(1)}$ and $\cL^{(2)}$
is a better choice.

The proponent of the ChPT might assert that,
to a given chiral order,
$\cL^{(1)}$ is the unique choice (modulo field transformations)
and hence any other Lagrangians,
including $\cL^{(2)}$ based on PCAC,
that do not give the same physics as $\cL^{(1)}$
should be discarded.
This assertion is often condensed into
a statement: {\it There is nothing sacred about PCAC.}
However, the issue is more subtle$^{15}$.
As illustrated above, the difference between
$\cL^{(1)}$ and $\cL^{(2)}$ appears at the $\cO(\rho^2)$ level.
However, since $\cL^{(1)}$ is devoid of
terms containing $(\Nb N)^2$ such as the $\pi^2 (\Nb N)^2$ term
(which in the mean-field approximation would
give contributions of $\cO(\rho^2)$),
it goes beyond the accuracy of one's starting point
to discuss the difference of $\cO(\rho^2)$
between $\cL^{(1)}$ and $\cL^{(2)}$.
If $\cL^{(1)}$ were a fundamental Lagrangian,
one might still be able to justify the absence of terms
involving $(\Nb N)^n$ ($n \geq 2$) in eq.(\ref{eQa}).
[For instance, the QED Lagrangian
is bilinear in fermion fields.]
However, since $\cL^{(1)}$ is an effective Lagrangian,
one cannot a priori exclude from $\cL^{(1)}$
multiple-fermion terms
that contain $(\Nb N)^n$ ($n \geq 2$).
We also note that
usual ChPT tests involving single baryons
place no constraints on these ``non-standard" terms.
Thus, there is no compelling reason
to prefer $\cL^{(1)}$ to $\cL^{(2)}$.

Meanwhile, from the current-algebra-plus-PCAC viewpoint,
one might claim that $\cL^{(2)}$ is a ``natural" choice,
and that $\cL^{(1)}$ is an approximate Lagrangian
obtained from $\cL^{(2)}$
by ignoring the $\cO(\rho^2)$ terms in eq.(\ref{eQl}).
However, this assertion is subject to the same criticism
as above, and therefore $\cL^{(2)}$ cannot be considered as
a better approximation than $\cL^{(1)}$.

These observations clearly indicate
that the true understanding of
the difference between $\mpi^*(1)$ and $\mpi^*(2)$
requires a consistent treatment of terms of $\cO(\rho^2)$
in the effective Lagrangian itself.
In other words,
the discrepancy between $\mpi^*(1)$ and $\mpi^*(2)$
represents the effects of
two (or more) -nucleon interaction terms
which have not been considered up to now.
This is a new type of matter effect.
Usually, matter effects of $\cO(\rho^2)$
such as the Lorentz-Lorenz-Ericson-Ericson effect,
the in-medium modifications of $g_A$, $m_N$ etc.,
are regarded as well-defined corrections
to the linear-density approximation.
However, our argument shows that
there exists a class of matter effects
which arise from higher-order density terms
in the effective Lagrangian.
Although the form of these extra terms can
vary for different choices of extrapolating fields,
this extrapolating-field dependence
does {\it not} affect $\mpi^*$,
if no truncation is introduced to the chiral effective
Lagrangian, and if $G_{\rho}(x;\varphi)$ is
calculated exactly.
However, when an approximation is introduced either
in $\cL$ or in $G_{\rho}(x;\varphi)$,
the resulting $\mpi^*$ can become dependent on
the interpolating field.

In the above we concentrated on the sigma term
in the pion sector.
{\it Essentially the same argument holds for the
contribution of the sigma term in the kaon sector.}
We add here a few comments
on the relation of the above argument to
the latest development
in the ChPT approach to kaon condensation.

The basic problems with the earlier ChPT calculations were:
(i) chiral-counting was not done consistently;
(ii) the $K$-$N$ scattering amplitudes
did not possess a correct energy dependence
to reproduce the scattering data.
Regarding problem (i),
a systematic ChPT calculation that respects
chiral-order counting
within the framework of the heavy-fermion formalism
has been carried out to tree order
by Brown, Lee, Rho and Thorsson$^{10}$,
and to one-loop order
by Lee, Jung, Min and Rho$^{11}$.
As far as the ordinary chiral counting in vacuum
is concerned, these calculations are complete up
to the stated chiral orders,
and it is to be noted that multiple-fermion terms do not feature
in these calculations.
This may seem to justify the absence
of multple-fermion terms in $\cL^{(1)}$,
but we must remember that,
because a finite-density system has
an additional scale parameter $\rho$,
chiral counting here can be significantly more complicated
than in vacuum.
This warning becomes particularly important
in applying ChPT to high-density matter.
Thus, it is crucially important
to check whether the contributions of
multiple-fermion terms are as suppressed as
the ordinary chiral counting would indicate.
Until this point is clarified,
there is no good reason to ignore $\cO(\rho^2)$ terms
in the starting Lagrangian.
Even if one adopts the working hypothesis
that the ordinary chiral counting can be applied to
a high-density system,
the inclusion of meson loops in a ChPT
calculation must, for consistency, be accompanied by
the inclusion of at least two-nucleon terms.
In this sense also,
$\cO(\rho^2)$ terms like $\pi^2 (\Nb N)^2$
should be retained in the Lagrangian.

Regarding (ii),
Lee et al.\ $\!\!^8$ considered the energy dependence
coming from the one-loop diagrams and the
resonance $\Lambda^*(1405)$,
and were able to reproduce reasonably well the existing data
on the s-wave $K$-$N$ scattering amplitude.
The pronounced energy dependence
in the s-wave $\bar{K}$-$N$ ($I=0$) scattering amplitude
was reproduced by adjusting the resonance parameters
pertaining to $\Lambda^*$.
However, the accuracy of
available experimental data
is not sufficient to test
the energy dependence arising from the loop diagrams.
Kaon condensation being sensitive to
the energy behavior of the $K$-$N$
amplitudes from threshold ($\omega= m_K$)
down towards $\omega=0$,
an important question is whether
this subthreshold energy behavior is reproduced satisfactorily
by the one-loop corrections.
In the language of the empirical low-energy expansion,
\beq
T_{KN} = a + b (\omega^2 - m_K^2)
 + \cO ( (\omega^2 - m_K^2)^2) ,\lab{eQn}
\eeq
this means that the parameters in $\cL$ must
reproduce not only the s-wave $K$-$N$ scattering length $a$
but also the s-wave effective range $b$.
Unfortunately, the quality of available experimental data
does not allow us to carry out this program.
In the phenomenological method of ref.\ 17,
this difficulty is reflected in the fact
that the $\Sigma_{KN}$ had to be treated
as a free parameter.
Further experimental information
on low-energy $K$-$N$ scattering as well as
a systematic calculation that includes $\cO(\rho^2)$ terms
are needed to make progress in this problem.

\vspace*{1cm}
I now move on to my second topic and discuss
the non-relativistic approximation
$<\!\!\bar{\psi}\psi\!\!>\,\sim\,
<\!\!\bar{\psi}\gamma_0\psi\!\!>$.
Again, to highlight my main point, I will concentrate on
the sigma term contribution.
Let us recall that a term in the effective Lagrangian
that engenders the sigma term is of the form
$m\bar{\psi}\psi$, where $\psi$ is either the quark or
baryon field, depending on whether one is working
at the quark level or at the hadronic level, and correspondingly
$m$ is either the quark or baryon mass.
Then, in the mean field approximation,
the Lorenz scalar density
$\rho_s \equiv \,<\!\!\bar{\psi}\psi\!\!>$
rather than the baryon density
$\rho \equiv\,<\!\!\bar{\psi}\gamma_0\psi\!\!>$
should measure the size of the $\sigma$ term contribution
in baryonic matter.
In the study of kaon condensation,
however, one has always been using
the approximation $\rho_s \sim \rho$.
The relatively large nucleon mass (at least, its free-space value)
motivates this approximation.
Another motivation may be that
$\rho$ is a conserved quantity that can be specified
as an external parameter whereas $\rho_s$ is a dynamical
quantity whose value can be obtained
only by solving the dynamics of the system.
As a caveat against this practice, however,
one should be reminded that,
in the relativistic mean field theory$^{21}$,
the distinction between $\rho_s$ and $\rho$
plays a crucial role.

A qualitative estimate
of the consequence of distinguishing between $\rho_s$ and $\rho$
may be obtained as follows.
For a nucleon of effective mass $\mN^*$ and momentum $\bfk$,
\ben
\bar{u}(\bfk)u(\bfk)\,=\,\frac{\mN^*}{E^*(\bfk)}
u^\dagger(\bfk)u(\bfk)\,,\;\;\;\;\;\;
E^*(\bfk)\equiv\sqrt{\mN^{*2}+\bfk^2}
\een
so that using $\rho$ instead of $\rho_s$
would overestimate the contribution of the sigma term
by a factor of $\kappa\,\equiv\,<\!E^*(\bfk)/\mN^*\!>$,
where $<\!\cdots\!>$
stands for averaging over the Fermi sea.
For $\mN^*=\mN$, we have $\kappa\;\sim 1.02$
at $\rho=\rho_0$ and $\kappa\;\sim 1.05$
at $\rho=3\rho_0$, but
the effect will become much more enhanced
as $\mN^*$ diminishes.

As a matter of fact,
some of the early works on pion condensation
paid due attention to the relativistic effect.
For example, in a classic paper
by Campbell, Dashen and Manassah$^{20}$,
the effective single-particle energy of a nucleon
was evaluated by applying an appropriate
Foldy-Wouthuysen transformation
to a relativistic effective Hamiltonian in the presence of
a background pion field.
A systematic elimination of ``odd"-operators ensured
the inclusion of relativistic effects to the relevant order.
A subsequent development on kaon condensation
did not quite follow the example
of ref.\ 20.
I wish to present here an exploratory study
of this relativistic effect.

For the illustrative purpose,
I again consider s-wave pion condensation
as a prototype of kaon condensation,
and adapt Campbell et al.\ $\!\!$'s treatment$^{20}$
of ($p$-wave) condensation based
on the linear $\sigma$ model.
A similar formulation is possible for
kaon condensation with the use of
the ``V-spin sigma model"$^{3,12}$.

Our Hamiltonian,
with all inessential terms dropped, is given by$^{12}$

\newpage
\ben
\cH &=& \half\,[(\nabla\sigma)^2+p_{\sigma}^2
  +(\nabla\bfpi)^2+{\bf p}_{\pi}^2]
+\,\lambda\,(\bfpi^2\,+\,\sigma^2\,-\,f^2)^2
-fm_{\pi}^2\sigma\non\\
&& +\,\psibarN[-i\bfgam\!\cdot\!\nabla
       +g\,(\sigma+i\gamma_5\,\bftau\!\cdot\!\bfpi)\,]\psiN
+\eta\,\psibarN\psiN\;.\lab{eq:Hsigma}
\een
Here $p_\sigma$ and ${\bf p}_{\pi}$ are the conjugate momenta,
and $\psiN$ is the nucleon doublet ($p,n$),
and $\eta\psibarN\psiN$
represents a baryon-density-dependent
chiral-symmetry breaking term.
The ground state of $\cH$ when the
baryon number is zero is characterized by $<\!\!\sigma\!\!>=f$, and
$<\!\!\bfpi\!\!>=0$.  (A slight deviation from $<\!\!\sigma\!\!>=f$
due to the symmetry breaking term
$-fm_{\pi}^2\sigma\non$ is ignored here.)
Thus, the vacuum is a $\sigma$-condensed state.
We now assume that for a sufficiently large baryon density,
the system develops a finite ground-state expectation value
for the pion.
For the sake of definiteness, let us consider
$\pi^0$-condensation, i.e., $<\!\!\pi^0\!\!>\neq 0$.
The s-wave condensation implies that
we can assume the meson condensate to be spatially uniform.
In this particular case $\cH$ becomes
\ben
\cH &=&-fm_{\pi}^2\sigma
+\lambda\,[\,(\pi^0)^2+\sigma^2-f^2]^2\non\\
& &+\,\psibarN[-i\bfgam\!\cdot\!\nabla
       +\,g\,(\sigma+i\gamma_5\,\tau_3\,\pi^0)\,]\psiN
+\,\eta\,\psibarN\psiN\;.\lab{eq:Hsigma'}
\een
Since the energy surface
in the $\sigma$-$\pi^0$ plane has a
sharp minimum along the ``magic" circle:
$\sigma^2+(\pi^0)^2=f^2$,
we can expect that energy minimum will shift
from its vacuum position
[$<\!\!\sigma\!\!>=f$, $<\!\!\pi^0\!\!>=0$]
along this magic circle.
We parametrize its new location as
\beq
<\!\pi^0\!>=f\sin\th\;,\;\;\;\;<\!\sigma\!>=f\cos\th.
\eeq
Then $\cH$ becomes
\beq
\cH=-f^2m_{\pi}^2\cos\th +\cH_N\;,\lab{eq:Hreduced}
\eeq
where the nucleon part $\cH_N$ is
\beq
\cH_N=\psibarN[-i\bfgam\!\cdot\!\nabla
       +g\,f\,(\cos\th+i\gamma_5\,\tau_3\,\sin\th)]\psiN
+\eta\psibarN\psiN
\eeq
For a single nucleon in free space, one should have
\beq
\cH_N(\th=0)=\psibarN
[-i\bfgam\!\cdot\!\nabla+(gf+\eta)]\psiN
\equiv\psibarN[-i\bfgam\!\cdot\!\nabla+m_N]\psiN,
\eeq
so that
\beq
gf\;=\;\mN - \eta \;\equiv \;\tilde{m} \lab{eq:mtilde}
\eeq
Then
\beq
\cH_N=\psibarN[-i\bfgam\!\cdot\!\nabla
       +\tilde{m}(\cos\th+i\gamma_5\tau_3\th)]\psiN
+\eta\psibarN\psiN
\eeq
To study the threshold behavior,
we consider a small-amplitude case ($\th\ll 1$),
which simplifies the above expression to
\beq
\cH_N=\psibarN[-i\bfgam\!\cdot\!\nabla \,+\,\mN +
\tilde{m}(-\frac{\th^2}{2}+i\gamma_5\,\tau_3\,\th)\,]\,\psiN
\eeq
As it stands, $\cH_N$ involves an ``odd" operator
accompanied by the large mass $\tilde{m}$.
With an appropriate Foldy-Wouthuysen transformation
$\cH_N$ becomes
\beq
\cH_N=\beta\,\mN+\beta\,\tilde{m}\,(-\frac{\th^2}{2})
+\frac{\beta}{2\mN}(\bfp\!\cdot\!
\mbox{\bm $\alpha$}
+i\tilde{m}\,\beta\,\gamma_5\,\tau_3\,\th)^2\,,
\eeq
which gives the single-particle energy $\epsilon(\bfp)$
of a plane-wave state with momentum $\bfp$ as
\beq
\epsilon(\bfp)\,=\,\mN+\tilde{m}\,(-\frac{\th^2}{2})
+\frac{1}{2\mN}(\bfp^2+\tilde{m}^2\th^2)
\eeq
The total nucleon energy $E_N$ is obtained by summing
$\epsilon(\bfp)$
up to the Fermi surface:
\beq
E_N=\int^{p<p_F}\!d^3p\;\epsilon(\bfp)
=E_0- \eta\;\frac{\tilde{m}}{\mN}\;
\frac{\th^2}{2}\;\rho,
\eeq
where $E_0$ represents the $\th$-independent contribution.
Adding the meson energy part, we obtain the total energy
\ben
E(\th)&=&-f^2 m_{\pi}^2\cos\th + E_N \noin\\
&\simeq& -f^2m_\pi^2\,(1-\frac{\th^2}{2})
+\eta\;\frac{\tilde{m}}{\mN}\;\frac{\th^2}{2}\,\rho.
\een
The critical density $\rho_c$ is a density at which
the coefficient of the $\th^2$ term becomes positive so that
$\th \ne 0$ gives lower energy than $\th=0$;
thus
\beq
\rho_c=\frac{f^2 m_{\pi}^2}{\eta}\!\cdot\!
\frac{\mN}{\tilde{m}} \lab{eq:rhocnew}
\eeq

The result eq.(\ref{eq:rhocnew}),
with $\eta=\sigma_{\pi N}$ understood,
should be compared
with eq.(\ref{eQd}), which was obtained
with the use of the non-relativistic approximation
$<\!\!\psibarN\psiN\!\!>\equiv\rho_s\ra\rho$.
Since $\mN/\tilde{m}=\mN/(\mN-\eta)>1$,
eq.(\ref{eq:rhocnew}) gives a higher value of $\rho_c$.
In the case of $s$-wave pion condensation,
$\eta\approx\sigma_{\pi N}\sim$ 50 MeV, so
the relativistic effect is not very important.
However, in a similar calculation for the kaon sector$^3$,
$\eta\approx\sigma_{KN}\sim$ 500 MeV so that,
even for the free-space value of $m_N$,
$\mN/\tilde{m}\sim 2$, which is a sizable effect.
As $\mN$ is reduced to $\mN^*$ in medium,
the effect would be even more pronounced.

In fact, the authors of ref.\ 3 knew these features,
but they chose not to discuss the relativistic effect
because the accuracy of
the Lagrangian used in ref.\ 3
did not seem to warrant that.
One must be careful not to include arbitrarily
some of higher order effects.
It is my observation
that a subsequent remarkable development
in the treatment of higher order terms in chiral counting,
due to the Stony Brook-Saclay-Seoul group$^{10,11}$,
has now made it worthwhile to revisit this issue.
One thing to be emphasized here is that,
although the qualitative features of the
relativistic effects may be learned from specific models
(such as the above toy model and the relativisitc mean field
approach$^{21}$), a quantitative answer can
come only from a systematic in-medium chiral counting.
In this connection, it seems interesting to study
how the relativistic effects can be incorporated
into the heavy-baryon formalism$^{22}$.

\vspace*{0.8cm}
To summarize, I have discussed two major problems
we need to solve in order to draw
a firm conclusion on kaon condensation.
Solving either of them
requires a systematic chiral-counting formalism
applicable to dense many-body system.
This is a very exciting challenge because
any development along this line
will have many important ramifications
in addition to clarifying
the particular issue of kaon condensation.

\vspace{1.3cm}
\noin
{\it Acknowledgement}

A substantial part of this talk derives from a recent paper
by Hiroyuki Yabu, Fred Myhrer and myself$^{15}$,
and I wish to thank Hiro and Fred for their support
for this presentation.
I am deeply indebted to Mannque Rho
for many illuminating discussions.

I also would like to take this opportunity
to express my sincere gratitude to Professor Akito Arima,
who many years ago kindly accepted me as one of his
graduate students
and initiated me into the exciting field of nuclear physics.

\vspace{1.5cm}
\noin
{\bf References}
\newcounter{ncite}
\begin{list}{\arabic{ncite}}{\usecounter{ncite}}

\item
D.B. Kaplan and A.E. Nelson,
\PL {\bf B175} (1986) 57; {\bf B179} (1986) 409(E).

\item
A.E. Nelson and D.B. Kaplan, \PL {\bf B192} (1987) 193.

\item
G.E. Brown, K. Kubodera and M. Rho, \PL{\bf B192} (1987) 273.

\item
G.E. Brown, K. Kubodera, D. Page and P. Pizzochero,
\PR {\bf D37} (1988) 2042.

\item
T. Tatsumi, {\it Prog. Theor. Phys.} {\bf 80} (1988) 22.

\item
D. Page and E. Baron, {\it Astrophys. J.}\ {\bf 354} (1990) L17.

\item
H.D. Politzer and M.B. Wise, \PL{\bf B273} (1991) 156.

\item
T. Muto and T. Tatsumi, \PL{\bf B283} (1992) 165.

\item
G.E. Brown, K. Kubodera, M. Rho and V. Thorsson,
 \PL{\bf B291} (1992) 355.

\item
G.E. Brown, C.-H. Lee, M. Rho and V. Thorsson,
\NP {\bf A567} (1994) 937.

\item
C.-H. Lee, H. Jung, D.-P. Min and M. Rho,
\PL {\bf B326} (1994) 14.

\item
For review, see {\it e.g.\ }, K. Kubodera, {\it J. Korean Phys. Soc.}
{\bf 26} (1993) S171.

\item
G.E. Brown and H.A. Bethe, {\it Astrophys. J.},
{\bf 423} (1994)  659.

\item
J. Delorme, M. Ericson and T.E.O. Ericson,
\PL {\bf B291}(1992) 379.

\item
H. Yabu, F. Myhrer and K. Kubodera,
``Meson condensation in dense matter reexamined",
preprint USC(NT)-94-1, to be published in
{\it Phys. Rev. D}
{}.
\item
H. Yabu, S. Nakamura and K. Kubodera,
\PL {\bf B317}(1993) 269.

\item
H. Yabu, S. Nakamura, F. Myhrer and K. Kubodera,
\PL {\bf B315} (1993) 17.

\item
M. Lutz, A. Steiner and W. Weise,
\PL {\bf B278} (1992) 29.

\item
J. Gasser and H. Leutwyler, {\it Ann. Phys.}
{\bf 158} (1984) 142.

\item
D.K. Cambell, R.F. Dashen and J.T. Manassah,
\PR {\bf D12} (1975) 979.

\item
B.D. Serot and J.D. Walecka, in
{\it Advances in Nuclear Physics}, vol. 16,
eds. J.W. Negle and E. W. Vogt (Plenum, New York, 1985), p. 1.

\item
H. Georgi, \PL {\bf B240} (1990) 447;\\
E. Jenkins and A. Manohar, \PL {\bf B255} (1991) 558.

\end{list}

\end{document}